\documentclass[conference,a4paper]{IEEEtran}
\usepackage{ifthen}
\newboolean{templateconf}
\setboolean{templateconf}{true}
\usepackage{slashbox}

\usepackage[utf8]{inputenc}
\usepackage[T1]{fontenc}
\usepackage{microtype}
\usepackage{amsmath}
\usepackage{amssymb}

\usepackage{mathtools}

\DeclarePairedDelimiter\floor{\lfloor}{\rfloor}

\iftemplateconf
\else
\usepackage{amsthm}
\usepackage[numbers]{natbib}
\fi
\usepackage[english]{babel}
\usepackage[autostyle]{csquotes}
\usepackage[dvipsnames]{xcolor}
\usepackage{graphicx}
\graphicspath{{./}{./}{./}}
\usepackage[backgroundcolor=orange]{todonotes}
\usepackage{siunitx}
\usepackage{tikz}
\usetikzlibrary{arrows,automata,calc,fit,patterns,positioning,backgrounds,shadows,shapes,decorations}
\usepackage{relsize}
\usepackage{ctable}
\usepackage{multirow}
\usepackage{setspace}
\usepackage{paralist}
\usepackage[bookmarks=false]{hyperref}
\hypersetup{linkcolor=black,citecolor=black,urlcolor=black,colorlinks=true}

\addto\extrasenglish{%
}

\iftemplateconf
\IEEEoverridecommandlockouts
\title{\Large\textbf{High Precision Indoor Navigation for Autonomous Vehicles}}
\author{Eduardo Sánchez Morales$^{1}$, Michael Botsch$^{1}$, Bertold Huber$^{2}$ and Andrés García Higuera$^{3}$%
	\thanks{$^{1}$Eduardo Sánchez Morales and Michael Botsch are with the Department of Vehicle Safety, CARISSMA, Technische Hochschule Ingolstadt, Ingolstadt, Germany.}%
	\thanks{$^{2}$Bertold Huber is with GeneSys Elektronik GmbH, Offenburg, Germany.}%
	\thanks{$^{3}$Andrés García Higuera is with the Department of Electrical, Electronic and Automation Engineering, University of Castilla-La Mancha, Ciudad Real, Spain.}%
}
\else
\title{High Precision Indoor Navigation for Autonomous Vehicles}
\author{
	\IEEEauthorblockN{%
		Eduardo Sánchez Morales,\\
		Michael Botsch\\
	}\\
	\IEEEauthorblockA{%
		Department of Vehicle Safety\\
		CARISSMA\\
		Technische Hochschule Ingolstadt\\
		Ingolstadt, Germany
	}
}
\fi

\IEEEoverridecommandlockouts

\makeatletter
\let\old@ps@headings\ps@headings
\let\old@ps@IEEEtitlepagestyle\ps@IEEEtitlepagestyle
\def\confheader#1{%
	\def\ps@headings{%
		\old@ps@headings%
		\def\@oddhead{\strut\hfill#1\hfill\strut}%
		\def\@evenhead{\strut\hfill#1\hfill\strut}%
	}%
	\def\ps@IEEEtitlepagestyle{%
		\old@ps@IEEEtitlepagestyle%
		\def\@oddhead{\strut\hfill#1\hfill\strut}%
		\def\@evenhead{\strut\hfill#1\hfill\strut}%
	}%
	\ps@headings%
}
\makeatother

\begin{document}
	
	\bstctlcite{IEEEexample:BSTcontrol}
	\maketitle
	
	\tikzstyle{box}  = [shade,shading=axis,rectangle,draw,node distance=4mm and 4mm]%
\tikzstyle{boxb} = [box,rounded corners,top color=Cerulean,bottom color=MidnightBlue]%
\tikzstyle{boxg} = [box,rounded corners,top color=cyan,bottom color=cyan]%
\newcommand{\architecture}{%
    \node[boxg] (box0) {input};%
    \node[boxg] (box1) [right=of box0, align=center]{street data\\processing};%
    \node[boxg] (box2) [right=of box1, align=center]{trajectory\\generation};%
    \node[boxg] (box3) [right=of box2, align=center]{collision\\recognition};%
    \node[boxg] (box4) [right=of box3, align=center]{risk\\assessment};%
    \node[boxg] (box5) [right=of box4, align=center]{output};%
    \path[very thick,->] (box0) edge (box1);%
    \path[very thick,->] (box1) edge (box2);%
    \path[very thick,->] (box2) edge (box3);%
    \path[very thick,->] (box3) edge (box4);%
    \path[very thick,->] (box4) edge (box5);%
    \node[fit=(box1) (box4)] (box-cover) {};%
    \node[node distance=0mm and 0mm] (box-label) [above=of box-cover] {algorithm framework};%
    \begin{scope}[on background layer]
        \node[boxb, fit=(box-cover) (box-label)] (box-background) {};%
    \end{scope}
}%

	\begin{abstract}
		Autonomous driving is an important trend of the automotive industry. The continuous research towards this goal requires a precise reference vehicle state estimation under all circumstances in order to develop and test autonomous vehicle functions. However, even when lane-accurate positioning is expected from oncoming technologies, like the L5 GPS band, the question of accurate positioning in roofed areas, e.\,g., tunnels or park houses, still has to be addressed.
		
		In this paper, a novel procedure for a reference vehicle state estimation is presented. The procedure includes three main components. First, a robust standstill detection based purely on signals from an Inertial Measurement Unit. Second, a vehicle state estimation by means of statistical filtering. Third, a high accuracy LiDAR-based positioning method that delivers velocity, position and orientation correction data with a mean error of 0.1 m/s, 4.7 cm and 1$^\circ$ respectively. Runtime tests on a CPU indicates the possibility of real-time implementation.
	\end{abstract}
	
	\iftemplateconf
	\else
	\fi
	
	\section{Introduction and motivation}\label{sec:intro}
	
	Arguably, one of the most important trends in the automotive industry, is Autonomous Driving. Manufacturers like Tesla are already equipping their entry level vehicles with emergency braking, collision warning and blind spot monitoring, and offering other Advanced Driver Assistance Systems (ADAS) as options, like Autopilot, Auto Lane Change, Autopark and Summon~\cite{TeslaAutoPilot}. Even when these ADAS still expect a human in the loop, they are clearly pushing towards a level 5 vehicle automation~\cite{j3016}.
	
	On the other hand, to achieve the goal of full-automation for vehicles, a high precision and reliability under all circumstances is demanded from the vehicle sensors and functions. As with any other apparatus that can cause harm to humans because of malfunction, autonomous vehicles have to offer a safe operation under adverse circumstances, safety redundancies, fall-back and recovery mechanisms. This requires extensive testing of the vehicle sensors and autonomous functions under various environments, which demands adequate referencing.
	
	Continuous research on sensor technology is rapidly advancing to achieve the high precision and reliability required for the developing of autonomous driving functions. That is the case of consumer-grade Satellite Navigation (SatNav) receivers with lane-accurate positioning~\cite{BroadcomGPS}, the release of the L5 GPS band~\cite{GPSL5} and the continuous development of Beidou~\cite{unicoreBeidou}. Nevertheless, there are situations where this high precision sensors have limitations that have to be addressed, such as park houses, proximity of tall buildings, among others.
	
	In the present work, three aspects of a precise reference vehicle state estimation that show a significant improving potential from current state-of-the-art are addressed. Starting with an Inertial Measurement Unit (IMU)-based standstill recognition, going through an accurate vehicle state prediction and finishing with a method for delivering correction data in roofed areas. The proposed algorithm is intended to be a reference for the developing and testing of autonomous functions. Hence, an important self-limitation is that no on-board information from the vehicle is used. The resulting system can be used as a reference state estimation system for developing and testing autonomous vehicles in roofed areas like parking houses or enclosed test facilities.
	
	This paper makes the following contributions:
	
	\begin{inparaenum}[1)]
		It shows
		\item a novel, robust, purely IMU-based standstill recognition (\autoref{sec:SSC}),
		\item an Extended Kalman Filter (EKF) that allows a prolonged and accurate vehicle state estimation without correction data (\autoref{sec:KF}),
		\item a novel, robust, highly-accurate LiDAR-based Positioning Method (LbPM) (\autoref{sec:IP}), and
		\item it combines these three elements, while keeping real-time implementation possibilities.
	\end{inparaenum}
	
	This paper is structured in the following manner: first, a brief review on the state-of-the art is given. Then, the methodology for each implemented module is explained. Finally, the evaluation methodology and the results are shown.	
	\section{State-of-the-Art}\label{sec:RelWork}
	
	\subsection{Standstill Detection}\label{sec:RW:SSD}
	
	Historically, velocity sensors have not been designed for measuring a standstill, but for measuring velocity over ground~\cite{SpeedoSchulze}. However, they have evolved to such a precision that they are used for estimating a standstill (null velocity over ground) with great accuracy. Some of these sensors are based on transducers~\cite{WPTKistler}. Other devices are equipped with an optical sensor and a lamp that are pointed downwards for information acquisition by means of an optical grid method~\cite{articleCorrevit}. These last sensors are very precise, but tend to face difficulties with low-contrast surfaces, such as water, snow or ice.
	
	A standstill recognition \textit{per se} has been addressed by Robert Bosch GmbH~\cite{SSBosch}. The patented method consists on fixing a camera on a vehicle facing outwards. The algorithm attempts to locate one same object in two different frames, and to derive a standstill statement from this information. Nevertheless, cameras are known for a strong trade-off between accuracy and distance to the seen objects, which affects negatively the accuracy of the standstill recognition.
	
	\subsection{Inertial Navigation Systems}\label{sec:RW:ReferenceSystems}
	
	Two of the state-of-the-art Inertial Navigation Systems (INS) that are used as references are the RT4003~\cite{OxfordRT4003} and the ADMA-G-PRO+~\cite{ADMAGenesys}. Both systems combine IMUs with Real-Time Kinematic (RTK) data for obtaining centimetre-precise position estimations in open areas. The RT4003 uses servo-accelerometers and Microelectromechanical Systems (MEMS) gyroscopes. The ADMA-G-PRO+ uses servo-accelerometers and optical gyroscopes, which provide better stability over longer periods of time since they are less prone to sensor noise and drift.
	
	Both systems are able to overcome short SatNav outages by relying on their high-precision IMUs. Yet, there is a limitation even for high-end sensors. Disturbances common in many vehicle testing environments, such as engine vibrations, cause the quality of the IMU navigation to dilute over time. It is precisely in situations with extended SatNav interruptions where these INSs have their biggest improving potential.
	
	\subsection{Indoor Positioning}\label{sec:RW:IP}
	
	On the indoor positioning field, comprehensive surveys on related research can be found in ~\cite{IPSOverview1} and ~\cite{IPSOverview2}. The methods seen in these publications rely on the combination of diverse sensors and technologies, such as sonar, radar or WiFi; proprietary technologies like iBeacon~\cite{iBcn}, and so on.
	
	The best performing LiDAR-based positioning method found is presented in ~\cite{Ibisch}. It consists on installing an arrange of LiDAR sensors on a parking lot and classifying the points of the point cloud as \textit{active} or \textit{static}. From the active points, the wheels of vehicles were detected, from which vehicle size, pose (position and orientation) and velocity is derived. An accuracy with a mean error of 11.5 cm and standard deviation $\left(\sigma\right)$ of  $5.4\si{\centi\meter}$ is shown, but the outputs of the algorithm are compared to \enquote{human-labeled} ground-truth data.
	
	The literature research that has been carried out reveals the Active Bat~\cite{ActiveBat2}~\cite{ActiveBat3} as the best performing indoor positioning system. The system consists of an ultra-sonic sender and an array of receivers. The sender is a small, spherical arrange of ultra-sonic speakers pointed upwards. The receivers are located on the roof of the room where the system is installed. They must be located in a square array pattern, with separations of 1.2m between receivers. Arguably, the numerous receivers required for the array is the biggest disadvantage of this system. The system shows an accuracy with a mean error of 3 cm is shown, but the evaluation methodology is not clearly specified~\cite{ActiveBat1}.
	
	Other consumer grade indoor positioning systems are available on the market~\cite{indrs}~\cite{Infsoft}~\cite{Senion}. Their accuracy is sufficient for their intended purpose, but still not adequate for automotive positioning.
	
	\section{Mathematical preamble}\label{sec:PrevMath}
	
	The coordinate systems used in this work are explained in the following. The vehicles move on the Local Tangent Plane (LTP). It is defined similar to the East-North-Up (ENU) coordinate system. So, $x_{\text{LTP}}$ points east, $y_{\text{LTP}}$ north and $z_{\text{LTP}}$ upwards, with $\overrightarrow{z_{\text{LTP}}}=\overrightarrow{x_{\text{LTP}}}\times\overrightarrow{y_{\text{LTP}}}$ and an arbitrary origin $o_{\text{LTP}}$ on the surface of the earth. The Local Car Plane (LCP) is defined similar to the ISO8855:2011 norm. So, $x_{\text{LCP}}$ points to the hood, $y_{\text{LCP}}$ to the driver, $z_{\text{LCP}}$ upwards, with $\overrightarrow{z_{\text{LCP}}}=\overrightarrow{x_{\text{LCP}}}\times\overrightarrow{y_{\text{LCP}}}$ and origin $o_{\text{LCP}}$ at the Center of sprung Mass (CoM) of the car. For simplification purposes, the $x_{\text{LCP}}y_{\text{LCP}}$-plane is assumed to be parallel to the $x_{\text{LTP}}y_{\text{LTP}}$-plane. It is assumed that all sensors mounted on the vehicle measure in the LCP, unless otherwise specified.
	
	Vectors are represented in boldface and matrices in boldface, capital letters. All the units are given in the International System of Units (SI), unless otherwise specified.
	
	\section{Methodology}\label{sec:algorithm}
	
	The procedure has three steps. The first one is to detect if the vehicle is in standstill. The second one is to estimate the vehicle state when the car is moving. The third one is to deliver correction data for the state estimation. Each step is explained in the following.
	
	\subsection{Standstill Classifier}\label{sec:SSC}
	
	\subsubsection{Introduction and Motivation}\label{sec:SSC:Why}
	
	The first step of the procedure is to detect if the vehicle is standing still. The proposed Standstill Classifier (SSC) is based on observing the signals of a strapped down IMU in the vehicle. This implies a huge testing simplification, as IMUs are common in testing setups, and no other additional sensors are required. This stand-alone nature of the method means a complete independence from any other sensor, making it robust in complex environments, such as roofed areas, where no SatNav is available.
	
	The standstill classification is done by means of a Random Forest (RF) classifier. The use of a machine learning (ML) approach is justified by several reasons. The main one is that it is not possible to establish manual thresholds on the IMU signals for a robust standstill detection. Much less when considering a variety of vehicles. This happens because some driving situations are almost equal to a standstill. Two examples are cruising on a straight highway or driving at walking velocity. Even if the false-negatives and the huge effort for trying to find thresholds are ignored, false-positives might still appear, which is unacceptable. The used standstill classes are shown in the \autoref{tab:SSClasses}.
	
	\newcommand{\mc}[1]{\multirow{2}{*}{#1}}
	\ctable[mincapwidth=\columnwidth,
	label=tab:SSClasses,doinside=\relscale{0.94},
	caption={Standstill classification.}]{ccc}{}{\FL
		\backslashbox{True state}{Detected state} & Standstill & Motion\LL
		Standstill & True positive & False negative\NN
		Motion & False positive & True negative\LL
	}
	
	The training and classifying processes for the RF can be found in the literature~\cite{breiman2001random}. The specific details applicable to this work are explained in the following.
	
	\subsubsection{Dataset generation}\label{sec:SSC:Dataset}
	
	For the RF training, a 10-minute long dataset is used. The car accelerations $a_{x_{\text{LCP}}}$, $a_{y_{\text{LCP}}}$ and $a_{z_{\text{LCP}}}$, and rotation rates $\dot{\phi}_{\text{LCP}}$, $\dot{\theta}_{\text{LCP}}$ and $\dot{\psi}_{\text{LCP}}$ (roll, pitch and yaw) along the LCP axes are recorded. The classification is done based on raw data, i.e., without performing any correction of the IMU data. The dataset should be equally split between the following three driving modes.
	
	\begin{inparaenum}[a)]
		\item Standstill. The car is placed in standstill and ready to drive. Automatic transmissions should be in drive and standard transmissions can either be in neutral or in gear with pressed clutch. Except the engine and transmission, all live loads in the car are undesired. All dead loads are unimportant.
		
		\item Walking velocity. The car is let to roll on a straight line as slow as possible, avoiding contact with the pedals or the steering wheel. Passengers are to avoid sudden motions.
		
		\item Normal driving. The car is driven with varying accelerations, velocities and steering angles. The more diverse the vehicle motion is, the better for the dataset.
	\end{inparaenum}
	
	This process applies for all kinds of vehicles, regardless of their motor train. The dataset can be generated in shorter cycles of the three driving modes.
	
	\subsubsection{Feature generation}\label{sec:SSC:Feature}
	The used features are divided in two groups. The features of the first group are (1)$a_{x_{\text{LCP}}}^2+a_{y_{\text{LCP}}}^2$, (2)$a_{z_{\text{LCP}}}^2$, (3)$\dot{\phi}_{\text{LCP}}^2+\dot{\theta}_{\text{LCP}}^2$ and (4)$\dot{\psi}_{\text{LCP}}^2$.
	
	From the features in the first group, a second group of features in the frequency domain is created. For this, a sliding window approach is used. So, sections of $\tau_{SSC}=170 \si{\milli\second}$ of the features 1-4 are taken, and a Discrete Fourier Transform (DFT) is generated from these sections. From the resulting frequency spectrum, the single-sided amplitude of some frequencies are taken as the next features. Considering the IMU sampling frequency $Fs=100 \si{\hertz}$, $N_\text{samples}=17$ samples are used for the DFT. As no DFT can be calculated for the first $N_\text{samples}-1$ samples, their single-sided amplitudes are set to zero. Because of this, it is recommended to start the dataset in standstill. Also, according to the Nyquist sampling rate, the maximum meaningful frequency from the DFT is $\frac{Fs}{2}=50 \si{\hertz}$. So, the used frequency range is described as
	\begin{align}
	\left[\frac{Fs}{N_\text{samples}}:\frac{Fs}{N_\text{samples}}:\floor[\Bigg]{\frac{N_\text{samples}}{2}}\cdot\frac{Fs}{N_\text{samples}}\right],
	\end{align}
	where $\floor[]{\,\,}$ is the floor operation.
	
\begin{figure}
		\centering
		\includegraphics[width=\columnwidth]{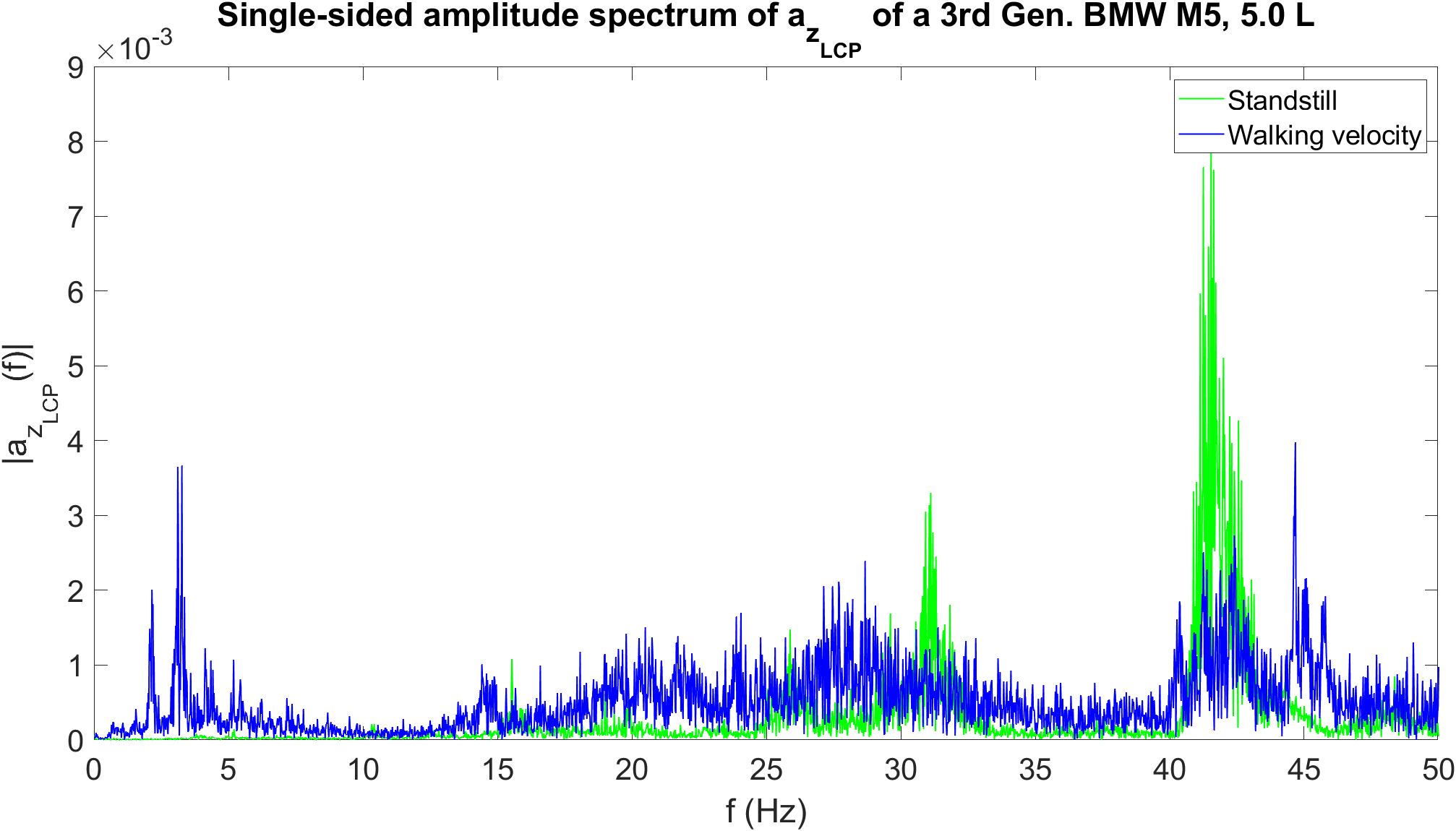}
		\caption{Single-sided amplitude spectrum of $a_{z_{\text{LCP}}}$ for a BMW M5}
		\label{fig:StandstillMotionFFTM5}
	\end{figure}

	The mentioned frequencies provide a robust classification for various vehicles, but different ones can be used for improving the classification performance of different cars. As shown in the \autoref{fig:StandstillMotionFFTM5}, some frequencies provide better classification for specific vehicles.
	
	From the above, 36 features for each sample are obtained: 4 in the time domain and 8 in the frequency domain for each of the first four features ($4+(4\cdot8)$). These features are chosen since they strongly correlate to the amount of kinetic energy of the vehicle.
	
	\subsubsection{Data labelling}\label{sec:SSC:label}
	
	For the labelling, state-of-the-art sensors can be taken as reference, provided they are not used under adverse conditions (SatNav in roofed areas or Correvit on ice, for example). The jerks that appear when the vehicle drives off and when it comes to a standstill, belong to the "motion" class. Remembering the mentioned sliding window, the used label is that of the $N_\text{samples}$-th sample. This means that drive-off and stopping transitions are labelled with the state the vehicle is going into.
	
	The used training parameters are: (i) number of trees: 12 and (ii) stopping criteria: minimum leaf size = 5.
	
	\subsection{Kalman filter}\label{sec:KF}
	
	The second step of the procedure is to predict the vehicle state when it is in motion. One of the most computationally efficient methods for estimating the \textit{optimum} state of a mobile object, assuming Markovian-Gaussian random processes, is the Kalman Filter (KF). The algorithm description can be found in the literature~\cite{Kalman1960}, and the details applicable to this work are described in the following. Since the used system model is non-linear, an EKF is used.
	
	\subsubsection{Relevant quantities in vector-matrix notation}\label{sec:KF:Matrices}
	
	The used state vector is defined as
	\begin{align}
	\boldsymbol{x}=&[x^{\text{LTP}}_{o_{\text{LCP}}},y^{\text{LTP}}_{o_{\text{LCP}}},\psi^{\text{LTP}}_{x_{\text{LCP}}},\dot{\psi}_{\text{LCP}},v^{\text{LTP}}_{o_{\text{LCP}}},\beta_{v,\text{LCP}}, \nonumber \\
	&\dot{\beta}_{v,\text{LCP}},\beta_{p,\text{LCP}},a_{x_{\text{LCP}}},a_{y_{\text{LCP}}},a_{y_{\text{circ,LCP}}}]^\text{T},
	\label{eqn:KFStateVector}
	\end{align}
	where $x^{\text{LTP}}_{o_{\text{LCP}}}$ and $y^{\text{LTP}}_{o_{\text{LCP}}}$ are the (x,y) position of $o_{\text{LCP}}$ in LTP, $\psi^{\text{LTP}}_{x_{\text{LCP}}}$ is the angle from $x_{\text{LTP}}$ to $x_{\text{LCP}}$ in LTP, $\dot{\psi}_{\text{LCP}}$ is the yaw rate around $z_{\text{LCP}}$, $v^{\text{LTP}}_{o_{\text{LCP}}}$ is the magnitude of the velocity over ground of $o_{\text{LCP}}$ in LTP, $\beta_{v,\text{LCP}}$ and $\beta_{p,\text{LCP}}$ are estimators of the sideslip angle of the bodywork of the vehicle, $\dot{\beta}_{v,\text{LCP}}$ is the rate of $\beta_{v,\text{LCP}}$, $a_{x_{\text{LCP}}}$ is the acceleration along the $x_{\text{LCP}}$ axis, and $a_{y_{\text{LCP}}}$ and $a_{y_{\text{circ,LCP}}}$ are acceleration estimators along the $y_{\text{LCP}}$ axis.
	
	Two state variables for the sideslip angle and two for the lateral acceleration are used to improve the performance of the EKF. The first sideslip estimator $\beta_{v,\text{LCP}}$ is based on the balance of moments of inertia and provides better performance when used for estimating $v^{\text{LTP}}_{o_{\text{LCP}}}$. Complementary, $\beta_{p,\text{LCP}}$ is based on a vehicle geometric model and offers better accuracy for estimating $x^{\text{LTP}}_{o_{\text{LCP}}}$ and $y^{\text{LTP}}_{o_{\text{LCP}}}$. For the lateral acceleration, $a_{y_{\text{LCP}}}$ is a low-pass filtered version of the IMU signal, while $a_{y_{\text{circ,LCP}}}$ assumes constant circular movement $\left(\beta_{p,\text{LCP}}=0\right)$, and aids in the plausibilization of $\boldsymbol{x}$, since $\beta_{p,\text{LCP}}\approx0$ and $a_{y_{\text{LCP}}}\approx a_{y_{\text{circ,LCP}}}$ when the vehicle is in tractive driving (i.e. not drifting)~\cite{DrifitingDynamics}. More detailed information about sideslip estimators can be found in~\cite{schramm2010modellbildung}.
	
	The measurement vector is defined as
	\begin{align}
	\boldsymbol{z_{\text{in}}}=&[x^{\text{LTP}}_{\text{in}},y^{\text{LTP}}_{\text{in}},\psi^{\text{LTP}}_{\text{in}},\dot{\psi}_{\text{LCP,in}},v^{\text{LTP}}_{\text{in}},\nonumber \\&\beta_{v,\text{LCP},\text{in}},a_{x_{\text{LCP},\text{in}}},a_{y_{\text{LCP},\text{in}}}]^\text{T},
	\label{eqn:KFMeasVector}
	\end{align}	
	where $x^{\text{LTP}}_{\text{in}}$ and $y^{\text{LTP}}_{\text{in}}$ are the (x,y) position of $o_{\text{LCP}}$ in LTP, $\psi^{\text{LTP}}_{\text{in}}$ is the angle from $x_{\text{LTP}}$ to $x_{\text{LCP}}$ in LTP, $\dot{\psi}_{\text{LCP,in}}$ is the yaw rate around $z_{\text{LCP}}$, $v^{\text{LTP}}_{\text{in}}$ is the magnitude of the velocity over ground of $o_{\text{LCP}}$ in LTP, $\beta_{v,\text{LCP},\text{in}}$ is the sideslip angle of the bodywork of the vehicle, $a_{x_{\text{LCP}}}$ is the acceleration along the $x_{\text{LCP}}$ axis, and $a_{y_{\text{LCP},\text{in}}}$ is the acceleration along the $y_{\text{LCP}}$ axis.
	The used measurement noise covariance matrix is defined as
	\begin{align}
	\boldsymbol{C_{\text{in}}}=\text{diag}(&c^2_{x^{\text{LTP}}_{\text{in}}},c^2_{y^{\text{LTP}}_{\text{in}}},c^2_{\psi^{\text{LTP}}_{\text{in}}},c^2_{\dot{\psi}_{\text{LCP,in}}},c^2_{v^{\text{LTP}}_{\text{in}}},\nonumber\\
	&c^2_{\beta_{v,\text{LCP},\text{in}}},c^2_{a_{x_{\text{LCP},\text{in}}}},c^2_{a_{y_{\text{LCP},\text{in}}}}).
	\label{eqn:KFsigmaMat}
	\end{align}
	
	\subsubsection{Plausibilization of measurement vector}\label{sec:KF:LowPass}
\begin{figure}
		\centering
		\includegraphics[width=\columnwidth]{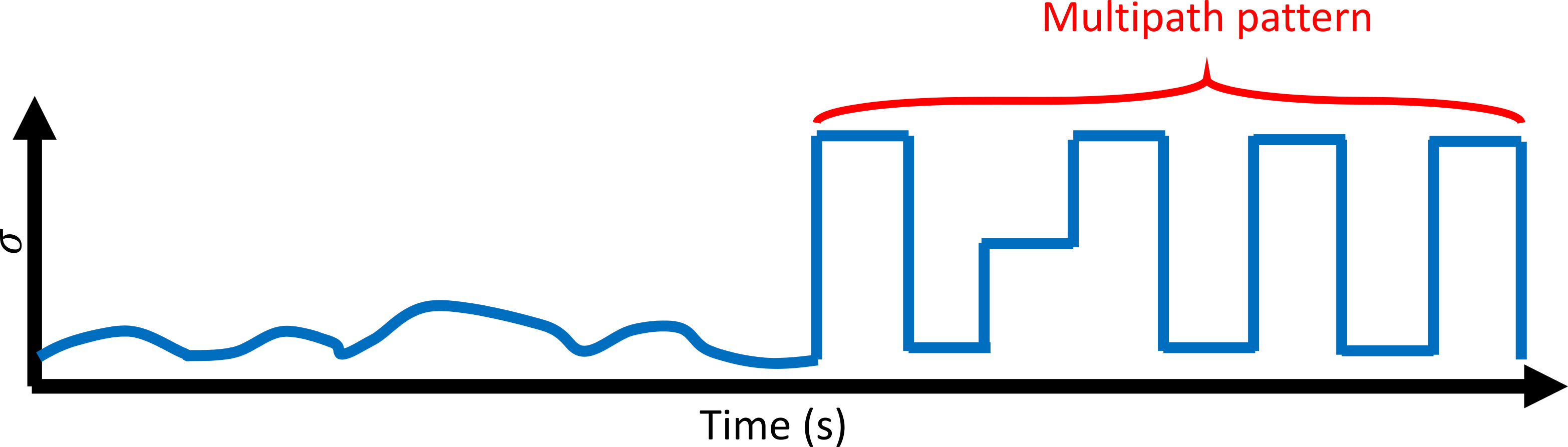}
		\caption{Qualitative illustration of a standard deviation signal affected by multipath.}
		\label{fig:mpp}
	\end{figure}

	An important limitation of the SatNav is the multipath effect that appears e.\,g.. in urban canyons. This causes the quality of the information to be degraded, and to $\sigma$ misjudgements from receivers that deliver this information. Patterns like the one shown in the \autoref{fig:mpp} are not uncommon. This affects greatly methods that rely on the $\sigma$ for sensor fusion, such as the EKF.	
	To address this, the $\sigma$ is dampened over time analogue to a PT1 element. For this, let $\sigma_{i_{\tau_{1}}}$ and $\sigma_{i_{\tau_{2}}}$ be the $\sigma$ for the $i$-th element of $\boldsymbol{z_{\text{in}}}$ provided by a SatNav receiver at time instances $\tau_{1}$ and $\tau_{2}$; $c_{i_{\tau_{1}}}$ and $c_{i_{\tau_{2}}}$ the $(i,i)$-th element of $\boldsymbol{C_{\text{in}}}$ at time instances $\tau_{1}$ and $\tau_{2}$; $T_{\tau_1}$ and $T_{\tau_2}$ the \textit{improving} time at time instances $\tau_{1}$ and $\tau_{2}$, $T_{\text{sat}}=1.5\,\si{\second}$ a saturation parameter, and $\Delta\tau=\tau_{2}-\tau_{1}$. Then, $T_{\tau_2}$ and $c_{i_{\tau_{2}}}$ are
	\begin{align}
	T_{\tau_2}=
	\begin{cases}
	0,&c_{i_{\tau_{1}}}<\sigma_{i_{\tau_{2}}}\\
	T_{\tau_1}+\Delta\tau,&\text{otherwise}
	\end{cases}, \text{and}
	\label{eqn:LowPassTime}
	\end{align}
	\begin{align}
	c_{i_{\tau_{2}}}\!\!=
	\begin{cases}
	\sigma_{i_{\tau_{2}}},\!\!\!\!\!\!\!\!\!\!\!\!\!\!\!\!\!\!&c_{i_{\tau_{1}}}<\sigma_{i_{\tau_{2}}}\!\!\!\!\\
	\!c_{i_{\tau_{1}}}e^{-\frac{T_{\tau_2}}{T_{\text{sat}}}}\!+\!\sigma_{i_{\tau_{2}}}\left(\!\!1\!-\!e^{-\frac{T_{\tau_2}}{T_{\text{sat}}}}\right)\!,&\text{otherwise.}\!\!\!\!
	\end{cases}
	\label{eqn:LowPassSigma}
	\end{align}
	The saturation $T_{\text{sat}}$ is optimized for the used SatNav receiver by recording the $\sigma$ patterns of urban canyons. The obtained 0.14:0.86 ratio for $c_{i_{\tau_{1}}}:\sigma_{i_{\tau_{2}}}$ at $T_{\tau_2}=3\si{\second}$ provides quick responses while filtering undesired $\sigma$ patterns.
	
	\subsection{LiDAR-based Positioning Method}\label{sec:IP}
	
	\subsubsection{Introduction and Motivation}\label{sec:IP:Why}
	The third step of the procedure is to deliver correction data for the predicted vehicle state. SatNav, one of the best sources of correction data, is not always available (\autoref{sec:intro}). Also, as literature research shows (\autoref{sec:RW:IP}), current indoor positioning methods might not meet the requirements (high accuracy) for autonomous vehicles or automotive safety research.
	
	The proposed LiDAR-based Positioning Method (LbPM) is based on using measurements from a mechanical LiDAR~\cite{HDL32Manual} to estimate the motion of a vehicle relative to fixed, precisely measured infrastructure points. From this relative motion, the velocity over ground, position and orientation are derived. The process is explained in the following.
	
	\subsubsection{Infrastructure markers}\label{sec:IP:Markers}
	
	Some infrastructure markers are used as reference points. The tape used in this work as marker~\cite{3mBand} complies with the UN/ECE 104 norm~\cite{ECE104}, which regulates retro-reflectors for trucks and trailers in Europe. However, existing infrastructural elements can be used as markers as well, provided that they are high-reflective, as is usual for road signs. The reflectivity is a critical aspect for the correct functioning of this method (\autoref{sec:IP:PointCloud}). 
	
	Since the markers are used as references, they are measured as precise as possible. Given that the position accuracy of the markers affects directly the performance of the LbPM, this could be considered a limitation. However, the precise location of the markers can be acquired with a Tachymeter, for example, in less than one hour for the test setup used in this work. This preparation time is feasible for many applications, as the one considered in this work.  An ID is assigned to each marker and a library containing the IDs and corresponding positions is saved for future use.
	
	The marker density affects mainly the frequency with which the correction data can be estimated. The more markers, the higher the probability of seeing one, thus a higher rate in the correction data estimation. As guidelines for the separation, the maximum distance to which a marker is seen with the desired reflectivity (\autoref{sec:IP:PointCloud}) in the performed tests is $d_{\text{velo,max}}\approx16$m. Differently, an extremely high marker density can lead to confuse one marker with another (\autoref{sec:IP:PointCloud}).
	
	\subsubsection{Point cloud managing}\label{sec:IP:PointCloud}
	
	The used LiDAR can measure the National Institute of Standards and Technology (NIST) calibrated reflectivity~\cite{NISTCal}. With this information, and given that the aim is to make the algorithm execute as fast as possible, the point cloud is first filtered with a NIST-reflectivity threshold $\Gamma_{\iota_{\text{velo}}}=200$. This filters out non-reflecting objects, such as walls, pillars, floor, and some mobile objects, greatly reducing the number of points to work with.
	
	Being a very efficient method that can process the points as the LiDAR packets arrive to the computer, the points are clustered by their timestamp with a threshold $\Gamma_{\tau_{\text{c,velo}}}=0.5\si{\milli\second}$. Taking the fastest rotation velocity of the LiDAR $\omega_{\text{velo,max}}=1200$RPM, the maximum horizontal Euclidean distance for point clustering is $d_\text{c,max}=d_{\text{velo,max}}\cdot\sin{\left(\omega_{\text{velo,max}}\cdot6\cdot \Gamma_{\tau_{\text{c,velo}}}\right)}=1$m. This is the theoretical minimum distance between markers to avoid mixing them up. The information of each cluster is then calculated using the information of the contained points and the mid-range arithmetic mean: $\frac{max-min}{2}$.
	
	\subsubsection{Marker identification}\label{sec:IP:MarkerID}
	
	The identification of the markers requires a rough car position $\boldsymbol{p}_\text{r}=[x_{\text{r}},y_{\text{r}}]^\text{T}$ 
	and a rough car orientation $\psi_\text{r}$ in LTP. Both can come from \textit{inaccurate} sensors, the state estimation from \autoref{sec:KF} or initial conditions at the beginning of a test. Assuming the true car position in LTP is given by $\boldsymbol{p}_\text{t}=[x_{\text{t}},y_{\text{t}}]^\text{T}$, $\boldsymbol{p}_\text{r}$ is constrained by $\left|\overrightarrow{\boldsymbol{p}_\text{t}\boldsymbol{p}_\text{r}}\right|<\frac{d_\text{c,max}}{2}$. Also, $\psi_\text{r}$ is constrained by $\left|\psi_\text{t}-\psi_\text{r}\right|<0.78\si{\radian}$.
	
	Knowing $\psi_\text{r}$, the LCP is oriented according to the LTP. Then, with a LiDAR measurement in LCP $\boldsymbol{p}_\text{m}=\left[x_{\text{m}},y_\text{m}\right]^\text{T}$, the apparent marker position in LTP is $\boldsymbol{p}_\text{s}=\boldsymbol{p}_\text{r}+\boldsymbol{p}_\text{m}$. This position is compared to those in the library (\autoref{sec:IP:Markers}). The marker closest to $\boldsymbol{p}_\text{s}$ is assumed to be the seen one. 
	
	\subsubsection{Velocity estimation}\label{sec:IP:Velocity}
\begin{figure}
		\centering
		\includegraphics[width=\columnwidth]{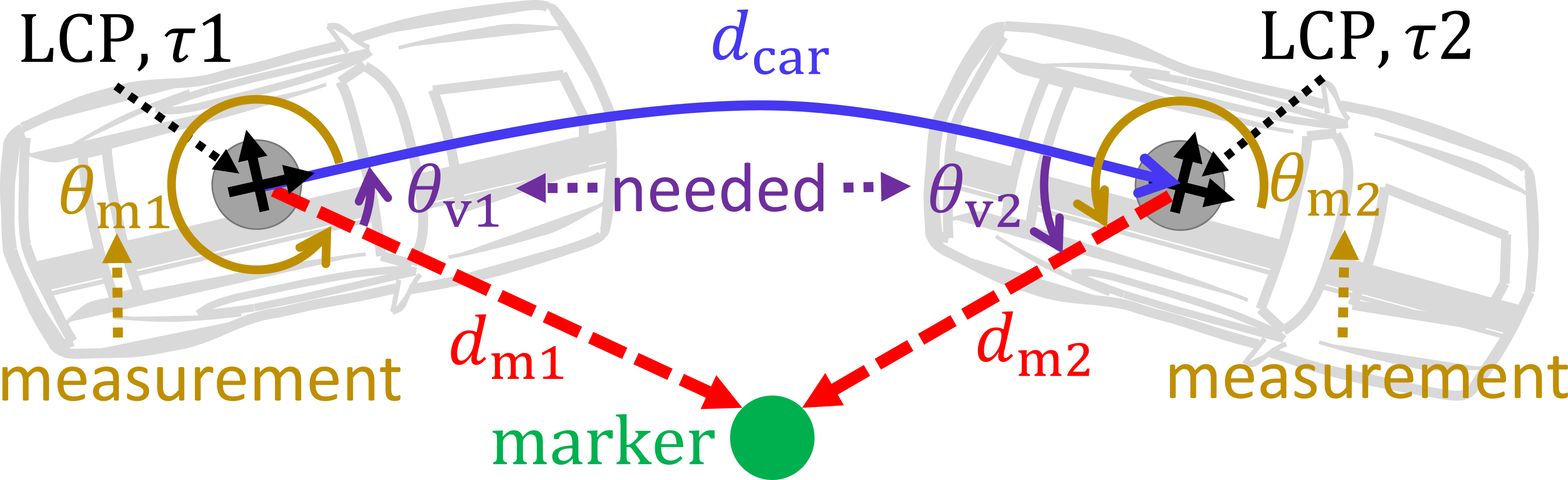}
		\caption{Illustration of the geometry for the velocity estimation.}
		\label{fig:DiagVel}
	\end{figure}

	For estimating the velocity, a cone shape is created from two LiDAR measurements $\boldsymbol{p}_\text{m1}=\left[x_\text{m1},y_\text{m1}\right]^\text{T}$ and $\boldsymbol{p}_\text{m2}=\left[x_\text{m2},y_\text{m2}\right]^\text{T}$ pointing to the same marker and made at time instances $\tau_{1}$ and $\tau_{2}$. So, let $d_\text{m1}$, $d_\text{m2}$, $\theta_\text{m1}$ and $\theta_\text{m2}$ be the distance and azimuth LiDAR measurements corresponding to $\boldsymbol{p}_\text{m1}$ and $\boldsymbol{p}_\text{m2}$. Given that the LCP can move relative to the LTP during $\Delta\tau$, the LCP at $\tau_1$ and $\tau_2$ is denoted with $\left(\text{LCP},\tau_{1}\right)$ and $\left(\text{LCP},\tau_{2}\right)$, respectively. So, the internal angles of the cone $\theta_\text{v1}$ and $\theta_\text{v2}$ are calculated as follows
	\begin{align}
	\theta_\text{v1}=
	\begin{cases}
	\theta_\text{m1},& \theta_\text{m1}<\pi\\
	2\pi-\theta_\text{m1},& \theta_\text{m1}>\pi
	\end{cases}, \text{and}
	\label{eqn:VeloVelo4}
	\end{align}
	\begin{align}
	\theta_\text{v2}=
	\begin{cases}
	\pi-\theta_\text{m2}+\Delta\tau\dot{\psi}_{\text{LCP}},& \theta_\text{m2}<\pi\\
	\theta_\text{m2}+\Delta\tau\dot{\psi}_{\text{LCP}}-\pi,& \theta_\text{m2}>\pi
	\end{cases}.
	\label{eqn:VeloVelo10}
	\end{align}
	Using the previous and the cosine law, the travelled distance during $\Delta\tau$ is given by 
	\begin{align}
	d_{\text{car}}\!\!=\!\!\sqrt{d_\text{m1}^2\!\!+\!d_\text{m2}^2\!\!-\!\!2 d_\text{m1}d_\text{m2}\!\cos(\pi\!-\!\theta_\text{v1}\!-\!\theta_\text{v2}\!+\!\Delta\tau\dot{\psi}_{\text{LCP}})}
	\label{eqn:TravelledDist}.
	\end{align}
	Assuming constant $v^{\text{LTP}}_{\text{in}}$ and $\dot{\psi}_{\text{LCP}}$ during $\Delta\tau$, the vehicle velocity over ground at $\tau_1$ and $\tau_2$ is $v^{\text{LTP}}_{\text{in}}=\frac{d_{\text{car}}}{\Delta\tau}$. A diagram of this process can be seen in the \autoref{fig:DiagVel}.
	\subsubsection{Pose estimation}\label{sec:IP:Position}
	For the pose estimation, $\boldsymbol{p}_\text{m1}$ and $\boldsymbol{p}_\text{m2}$ have to point to different markers. Assuming the car moves along an arch with constant radius, with constant $\dot{\psi}_{\text{LCP}}$ and $v^{\text{LTP}}_{\text{in}}$ during $\Delta\tau$, the radius of the arch is $r_{d_{\text{car}}}=\frac{v^{\text{LTP}}_{\text{in}}}{\dot{\psi}_{\text{LCP}}}$. So, the car movement during $\Delta\tau$  expressed in $\left(\text{LCP},\tau_{1}\right)$ is
	\begin{align}
	\begin{bmatrix}
	\Delta x_{\text{car}}\\
	\Delta y_{\text{car}}\\
	\Delta\psi_{\text{LCP}}
	\end{bmatrix}=
	\begin{bmatrix}
	r_{d_{\text{car}}}\cdot \sin(\Delta\tau\cdot\dot{\psi}_{\text{LCP}})\\
	r_{d_{\text{car}}}-\left(r_{d_{\text{car}}}\cdot \cos(\Delta\tau\cdot\dot{\psi}_{\text{LCP}})\right)\\
	\Delta\tau\cdot\dot{\psi}_{\text{LCP}}
	\end{bmatrix}.
	\label{eqn:PosVelo2}
	\end{align}
	Let the $\left(\text{LCP},\tau_{2}\right)$ be oriented according to the $\left(\text{LCP},\tau_{1}\right)$, and the position of two markers in LTP be $\boldsymbol{p}_\text{n1}=\left[x_\text{n1},y_\text{n1}\right]^\text{T}$ and $\boldsymbol{p}_\text{n2}=\left[x_\text{n2},y_\text{n2}\right]^\text{T}$. If $\boldsymbol{p}_\text{m1}$ and $\boldsymbol{p}_\text{m2}$ point to $\boldsymbol{p}_\text{n1}$ and $\boldsymbol{p}_\text{n2}$ respectively, then $\boldsymbol{p}_\text{n1}$ and $\boldsymbol{p}_\text{n2}$ are expressed in $\left(\text{LCP},\tau_{1}\right)$ as
	\begin{align}
	\begin{bmatrix}
	\boldsymbol{p}^{\tau_{1}}_{\text{n1}}\\
	\boldsymbol{p}^{\tau_{1}}_{\text{n2}}
	\end{bmatrix}=
	\begin{bmatrix}
	\boldsymbol{p}_\text{m1}\\
	\boldsymbol{p}_\text{m2}+\left[\Delta x_{\text{car}},\Delta y_{\text{car}}\right]^\text{T}
	\end{bmatrix}.
	\label{eqn:PosVelo999}
	\end{align}
	
\begin{figure}
		\centering
		\includegraphics[width=\columnwidth]{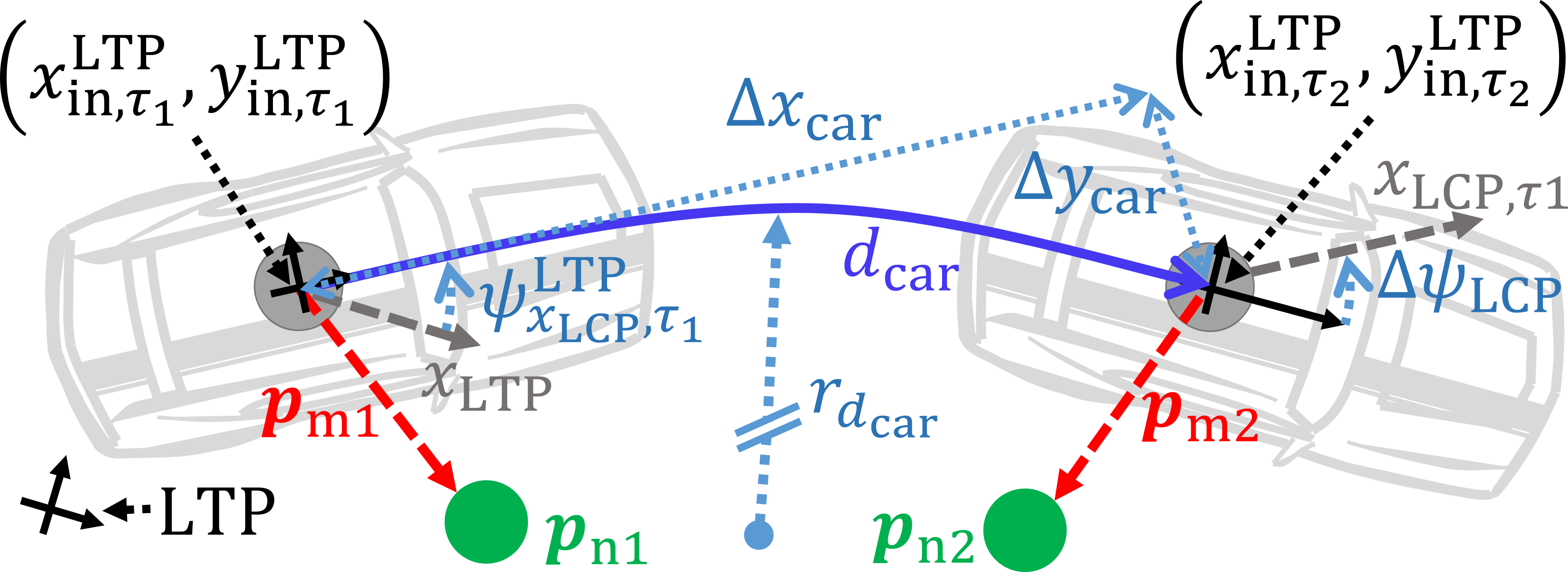}
		\caption{Illustration of the geometry for the position estimation.}
		\label{fig:DiagPos}
	\end{figure}

	Now, let $x_\text{n1}<x_\text{n2}$. The orientation of $\overrightarrow{\boldsymbol{p}_\text{n1}\boldsymbol{p}_\text{n2}}$ expressed in $\left(\text{LCP},\tau_{1}\right)$ is given by
	\begin{align}
	\theta_\text{n1n2}^{\tau_{1}}=
	\arctan 2\left(y_\text{n2}^{\tau_{1}}-y_\text{n1}^{\tau_{1}},x_\text{n2}^{\tau_{1}}-x_\text{n1}^{\tau_{1}}\right),
	\label{eqn:VectorEqSystVelo3}
	\end{align}
	and expressed in LTP is given by
	\begin{align}
	\theta_\text{n1n2}^{LTP}=
	\arctan 2\left(y_\text{n2}-y_\text{n1},x_\text{n2}-x_\text{n1}\right).
	\label{eqn:VectorEqSystVelo999}
	\end{align}
	The angular offset from $x_{\text{LTP}}$ to $x_{\text{LCP}}$ is by definition equal to $\psi^{\text{LTP}}_{x_{\text{LCP}}}$ (\autoref{sec:KF:Matrices}). So, at time instances $\tau_{1}$ and $\tau_{2}$ can be calculated as 
	\begin{align}
	\begin{bmatrix}
	\psi^{\text{LTP}}_{x_{\text{LCP}},\tau_{1}}\\
	\psi^{\text{LTP}}_{x_{\text{LCP}},\tau_{2}}
	\end{bmatrix}=
	\begin{bmatrix}\theta_\text{n1n2}^{LTP}-\theta_\text{n1n2}^{\tau_{1}}\\
	\theta_\text{n1n2}^{LTP}-\theta_\text{n1n2}^{\tau_{1}}+\Delta\psi_{\text{LCP}}
	\end{bmatrix}
	\label{eqn:psilcpfrommarker2}.
	\end{align}
	Since all measurements are now expressed in $\left(\text{LCP},\tau_{1}\right)$, $\psi^{\text{LTP}}_{x_{\text{LCP}},\tau_{1}}$ is used to rotate them. So,
	\begin{align}
	\begin{bmatrix}
	\boldsymbol{p}_{\text{n1}}^{\tau_{1},\psi^{\text{LTP}}_{x_{\text{LCP}},\tau_{1}}}\\
	\boldsymbol{p}_{\text{n2}}^{\tau_{1},\psi^{\text{LTP}}_{x_{\text{LCP}},\tau_{1}}}
	\end{bmatrix}=
	\begin{bmatrix}	
	\boldsymbol{R}\left(\psi^{\text{LTP}}_{x_{\text{LCP}},\tau_{1}}\right)\boldsymbol{p}_{\text{n1}}^{\tau_{1}}\\
	\boldsymbol{R}\left(\psi^{\text{LTP}}_{x_{\text{LCP}},\tau_{1}}\right)\boldsymbol{p}_{\text{n2}}^{\tau_{1}}
	\end{bmatrix}
	\label{eqn:rotatemarker2},
	\end{align}
	where $\boldsymbol{R}\left(\psi^{\text{LTP}}_{x_{\text{LCP}},\tau_{1}}\right)$ is a 2D rotation matrix of $\psi^{\text{LTP}}_{x_{\text{LCP}},\tau_{1}}$ radians around $z_{\text{LCP}}$. Finally, the linear offsets from $o_{\text{LTP}}$ to $o_{\text{LCP}}$ in LTP are by definition equal to $x^{\text{LTP}}_{\text{in}}$ and $y^{\text{LTP}}_{\text{in}}$ (\autoref{sec:KF:Matrices}). So, at time instance $\tau_{1}$ can be calculated as 
	\begin{align}
	\begin{bmatrix}
	x^{\text{LTP}}_{\text{in},\tau_{1}}\\
	y^{\text{LTP}}_{\text{in},\tau_{1}}
	\end{bmatrix}\!\!=\!\!\frac{\left(\!\boldsymbol{p}_\text{n1}\!-\!\boldsymbol{p}_{\text{n1}}^{\tau_{1},\psi^{\text{LTP}}_{x_{\text{LCP}},\tau_{1}}}\right)\!\!+\!\!\left(\!\boldsymbol{p}_\text{n2}\!-\!\boldsymbol{p}_{\text{n2}}^{\tau_{1},\psi^{\text{LTP}}_{x_{\text{LCP}},\tau_{1}}}\right)}{2}.
	\label{eqn:VectorEqSystVelo8}
	\end{align}
	Knowing $\Delta x_{\text{car}}$ and $\Delta y_{\text{car}}$, $x^{\text{LTP}}_{\text{in},\tau_{2}}$ and $y^{\text{LTP}}_{\text{in},\tau_{2}}$ are calculated as
	\begin{align}
	\begin{bmatrix}
	x^{\text{LTP}}_{\text{in},\tau_{2}}\\
	y^{\text{LTP}}_{\text{in},\tau_{2}}
	\end{bmatrix}=\begin{bmatrix}
	x^{\text{LTP}}_{\text{in},\tau_{1}}\\
	y^{\text{LTP}}_{\text{in},\tau_{1}}
	\end{bmatrix}+
	\begin{bmatrix}
	\Delta x_{\text{car}}\\
	\Delta y_{\text{car}}
	\end{bmatrix}
	\label{eqn:VectorEqSystVelo10}
	\end{align}
	A diagram of this process can be seen in the \autoref{fig:DiagPos}.
	
	\section{Evaluation and Results}\label{sec:results}
	
	For evaluating the algorithm modules, a state-of-the-art INS with RTK correction data is used outdoor. This device delivers accelerations and velocities in the LCP, and positions in the LTP. The SatNav-RTK correction data is used by the INS for refining all its outputs. An extensive and comprehensive test dataset of more than 8 hours of real-world driving is recorded with this INS and it includes city, freeway and highway driving; enclosed test facilities, open-sky roads, heavily wooded areas and multi-storey park houses; fluent, medium and heavy traffic conditions; gasoline, diesel and electric vehicles.
	
	\subsection{SSC Performance}\label{sec:SS:Performance}
	The performance of the SSC is tested in two ways. First, the classification is compared to the test dataset. A zero false-positive rate is obtained, and the standstill is robustly recognized when the vehicle comes to a stop.
	
	For the second performance test, the SSC is subjected to extreme cases. Two of these are explained following. 
	
\begin{figure}
		\centering
		\includegraphics[width=\columnwidth]{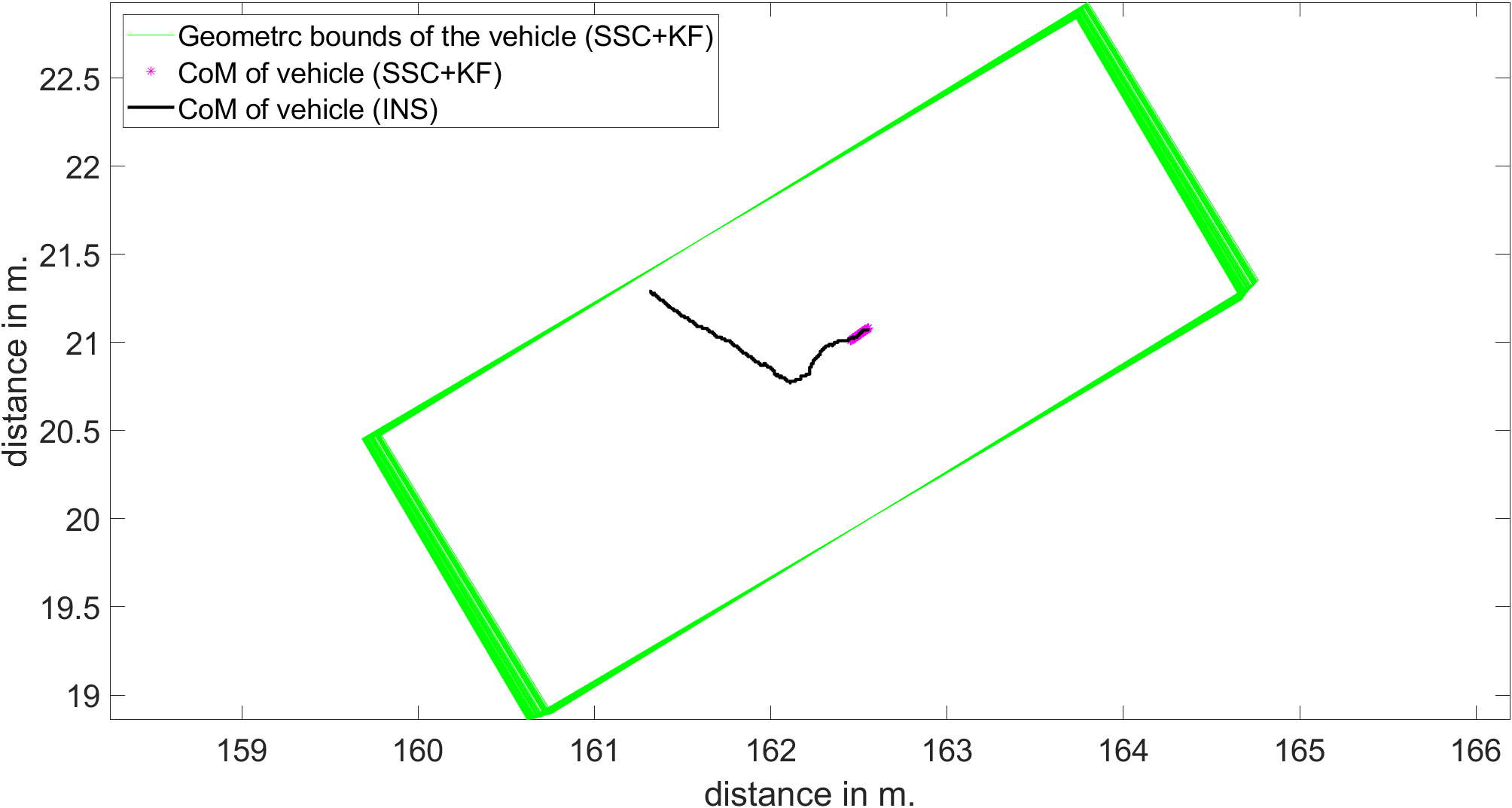}
		\caption{Comparison of estimated movement. SSC+EKF vs INS+RTK}
		\label{fig:SSQ7Map}
	\end{figure}

	In the first one, the vehicle is placed in standstill and high vibrations are induced continuously for 254s to its bodywork. The \autoref{fig:SSQ7Map} shows the results of this test case. The estimated movement by the proposed SSC+EKF method is 0.12m, while the INS estimated 1.26 m of motion. The reason is that the high, continuous vibrations, prevent the INS from recognizing that the vehicle is in standstill, even with RTK correction data. So, it keeps integrating the noise-induced IMU measurements that derive in velocities and apparent movement.
	
\begin{figure}
		\centering
		\includegraphics[width=\columnwidth]{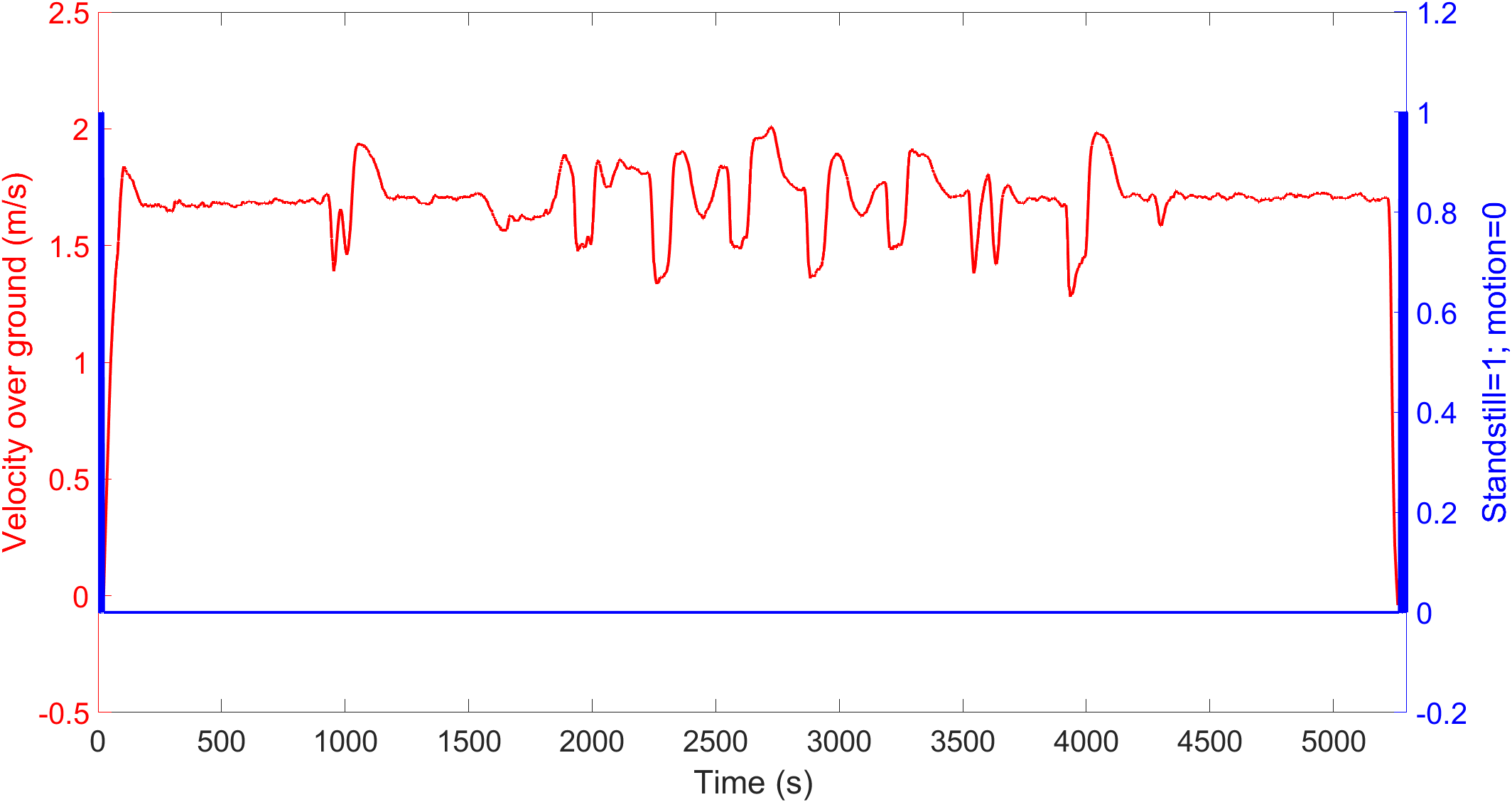}
		\caption{SSC and velocity over ground from an INS with RTK.}
		\label{fig:SSSmartSlow}
	\end{figure}

	For the second extreme case, a Smart Electric Drive is let to roll on a straight line. As seen in the results shown in the \autoref{fig:SSSmartSlow}, even with the absence of engine or transmission vibrations, and with minimal road influence, the SSC is able to recognize that the vehicle is in motion.
	
	\subsection{KF Performance}\label{sec:KF:Performance}
	For testing the performance of the EKF, the recorded position from the test dataset is compared to the position estimated by the proposed EKF. The reason is that the position is the magnitude that shows the biggest differences over time, when integrating accelerations and velocities. The recorded car accelerations along the LCP axes $a_{x_{\text{LCP},\text{in}}}$ and $a_{y_{\text{LCP},\text{in}}}$, and the yaw rate $\dot{\psi}_{\text{LCP,in}}$ are used in $\boldsymbol{z}_{\text{in}}$. No correction data is used.
	
	With a median velocity of $4.50 \frac{\si{\meter}}{\si{\second}}$ $\left(\approx16.21\frac{\si{\kilo\meter}}{\si{\hour}}\right)$, a median deviation of $254.94\si{\meter}$ per driven hour is obtained. 
	\subsection{LbPM Performance}\label{sec:LbPM:Performance}
	
	The LbPM is evaluated by comparing its outputs to those of the INS. For this, the LbPM is installed on an open-air test track for the sole purpose of having SatNav reception for the INS. It is important to note that the LbPM receives no correction data neither from the INS nor from the SatNav.
	
	For having a diverse dataset and to detect possible weaknesses of the LbPM, a drive-by and a slalom manoeuvrers are driven with varying velocities from 5 km/h and up to 40 km/h. The results of the \autoref{tab:LbPMVelresults}, \autoref{tab:LbPMPosresults} and \autoref{tab:LbPMYawresults} show that the obtained accuracy for velocity, position and orientation of the LbPM is similar to that of the INS-RTK reference. 
	
	A huge advantage of the proposed method, is that every iteration of the LbPM outputs the position, velocity and orientation for two time instances. At time instance $\tau_{2}$, the outputs for $\tau_{1}$ and $\tau_{2}$ are obtained. At $\tau_{3}$, the outputs for $\tau_{2}$ and $\tau_{3}$ are obtained, and so on. This allows to average two outputs corresponding to the same time instance, but made at different LbPM iterations. The effect of this is the compensation of measurement errors and the improving of the output accuracy. 
	
	\ctable[mincapwidth=\columnwidth,
	label=tab:LbPMVelresults,doinside=\relscale{0.94},
	caption={Velocity accuracy results for the proposed LbPM. Shown are the manoeuvrers, manoeuvrer velocity, mean deviation from INS, Std. dev. of the error and maximum deviation from INS.}]{cccc}{}{\FL
		\textbf{Drive-by} & \textbf{Mean $\left(\frac{\si{\meter}}{\si{\second}}\right)$}& \textbf{Std. dev. $\left(\frac{\si{\meter}}{\si{\second}}\right)$} & \textbf{Max. $\left(\frac{\si{\meter}}{\si{\second}}\right)$}\LL
		$5\frac{\si{\kilo\meter}}{\si{\hour}}$& 0.06 & 0.08 & 0.33 \NN
		$10\frac{\si{\kilo\meter}}{\si{\hour}}$& 0.08 & 0.10 & 0.57 \NN
		$15\frac{\si{\kilo\meter}}{\si{\hour}}$& 0.07 & 0.09 & 0.39 \NN
		$20\frac{\si{\kilo\meter}}{\si{\hour}}$& 0.08 & 0.09 & 0.50 \NN
		$25\frac{\si{\kilo\meter}}{\si{\hour}}$& 0.08 & 0.10 & 0.65 \NN
		$30\frac{\si{\kilo\meter}}{\si{\hour}}$& 0.08 & 0.10 & 0.57 \NN
		$35\frac{\si{\kilo\meter}}{\si{\hour}}$& 0.08 & 0.11 & 0.44 \NN
		$40\frac{\si{\kilo\meter}}{\si{\hour}}$& 0.11 & 0.13 & 0.47 \LL
		\textbf{Slalom} & \textbf{Mean $\left(\frac{\si{\meter}}{\si{\second}}\right)$}& \textbf{Std. dev. $\left(\frac{\si{\meter}}{\si{\second}}\right)$} & \textbf{Max. $\left(\frac{\si{\meter}}{\si{\second}}\right)$}\LL
		$5\frac{\si{\kilo\meter}}{\si{\hour}}$& 0.08 & 0.11 & 0.59 \NN
		$10\frac{\si{\kilo\meter}}{\si{\hour}}$& 0.09 & 0.12 & 0.59 \NN
		$20\frac{\si{\kilo\meter}}{\si{\hour}}$& 0.14 & 0.17 & 0.71 \NN
		$30\frac{\si{\kilo\meter}}{\si{\hour}}$& 0.18 & 0.24 & 0.76 \NN
		$40\frac{\si{\kilo\meter}}{\si{\hour}}$& 0.18 & 0.22 & 0.62 \LL
	}
	
	\ctable[mincapwidth=\columnwidth,
	label=tab:LbPMPosresults,doinside=\relscale{0.94},
	caption={Position accuracy results for the proposed LbPM. Shown are the manoeuvrers, manoeuvrer velocity, mean deviation from INS, Std. dev. of the error and maximum deviation from INS.}]{cccc}{}{\FL
		\textbf{Drive-by} & \textbf{Mean $\left(\si{\meter}\right)$}& \textbf{Std. dev. $\left(\si{\meter}\right)$} & \textbf{Max. $\left(\si{\meter}\right)$}\LL
		$5\frac{\si{\kilo\meter}}{\si{\hour}}$& 0.04 & 0.02 & 0.09 \NN
		$10\frac{\si{\kilo\meter}}{\si{\hour}}$& 0.03 & 0.02 & 0.10 \NN
		$15\frac{\si{\kilo\meter}}{\si{\hour}}$& 0.03 & 0.02 & 0.13 \NN
		$20\frac{\si{\kilo\meter}}{\si{\hour}}$& 0.03 & 0.02 & 0.09 \NN
		$25\frac{\si{\kilo\meter}}{\si{\hour}}$& 0.04 & 0.02 & 0.07 \NN
		$30\frac{\si{\kilo\meter}}{\si{\hour}}$& 0.06 & 0.02 & 0.10 \NN
		$35\frac{\si{\kilo\meter}}{\si{\hour}}$& 0.07 & 0.03 & 0.11 \NN
		$40\frac{\si{\kilo\meter}}{\si{\hour}}$& 0.08 & 0.03 & 0.15 \LL
		\textbf{Slalom} & \textbf{Mean $\left(\si{\meter}\right)$}& \textbf{Std. dev. $\left(\si{\meter}\right)$} & \textbf{Max. $\left(\si{\meter}\right)$}\LL
		$5\frac{\si{\kilo\meter}}{\si{\hour}}$& 0.04 & 0.02 & 0.12 \NN
		$10\frac{\si{\kilo\meter}}{\si{\hour}}$& 0.04 & 0.02 & 0.13 \NN
		$20\frac{\si{\kilo\meter}}{\si{\hour}}$& 0.04 & 0.02 & 0.10 \NN
		$30\frac{\si{\kilo\meter}}{\si{\hour}}$& 0.05 & 0.02 & 0.09 \NN
		$40\frac{\si{\kilo\meter}}{\si{\hour}}$& 0.10 & 0.02 & 0.12 \LL
	}
	
	\ctable[mincapwidth=\columnwidth,
	label=tab:LbPMYawresults,doinside=\relscale{0.94},
	caption={Orientation accuracy results for the proposed LbPM. Shown are the manoeuvrers, manoeuvrer velocity, mean deviation from INS, Std. dev. of the error and maximum deviation from INS.}]{cccc}{}{\FL
		\textbf{Drive-by} & \textbf{Mean $\left(\si{\deg}\right)$}& \textbf{Std. dev. $\left(\si{\deg}\right)$} & \textbf{Max. $\left(\si{\deg}\right)$}\LL
		$5\frac{\si{\kilo\meter}}{\si{\hour}}$& 0.73 & 0.25 & 1.48 \NN
		$10\frac{\si{\kilo\meter}}{\si{\hour}}$& 0.19 & 0.20 & 0.86 \NN
		$15\frac{\si{\kilo\meter}}{\si{\hour}}$& 0.26 & 0.19 & 0.69 \NN
		$20\frac{\si{\kilo\meter}}{\si{\hour}}$& 0.37 & 0.23 & 0.83 \NN
		$25\frac{\si{\kilo\meter}}{\si{\hour}}$& 0.58 & 0.23 & 0.84 \NN
		$30\frac{\si{\kilo\meter}}{\si{\hour}}$& 0.51 & 0.22 & 0.96 \NN
		$35\frac{\si{\kilo\meter}}{\si{\hour}}$& 0.44 & 0.25 & 0.88 \NN
		$40\frac{\si{\kilo\meter}}{\si{\hour}}$& 0.41 & 0.26 & 0.86 \LL
		\textbf{Slalom} & \textbf{Mean $\left(\si{\deg}\right)$}& \textbf{Std. dev. $\left(\si{\deg}\right)$} & \textbf{Max. $\left(\si{\deg}\right)$}\LL
		$5\frac{\si{\kilo\meter}}{\si{\hour}}$& 0.24 & 0.29 & 1.37 \NN
		$10\frac{\si{\kilo\meter}}{\si{\hour}}$& 0.40 & 0.29 & 1.22 \NN
		$20\frac{\si{\kilo\meter}}{\si{\hour}}$& 0.32 & 0.36 & 1.18 \NN
		$30\frac{\si{\kilo\meter}}{\si{\hour}}$& 0.36 & 0.40 & 1.28 \NN
		$40\frac{\si{\kilo\meter}}{\si{\hour}}$& 0.53 & 0.43 & 1.25 \LL
	}
	
	\subsection{Algorithm performance} \label{sec:Res:All}
	
\begin{figure}
		\centering
		\includegraphics[width=\columnwidth]{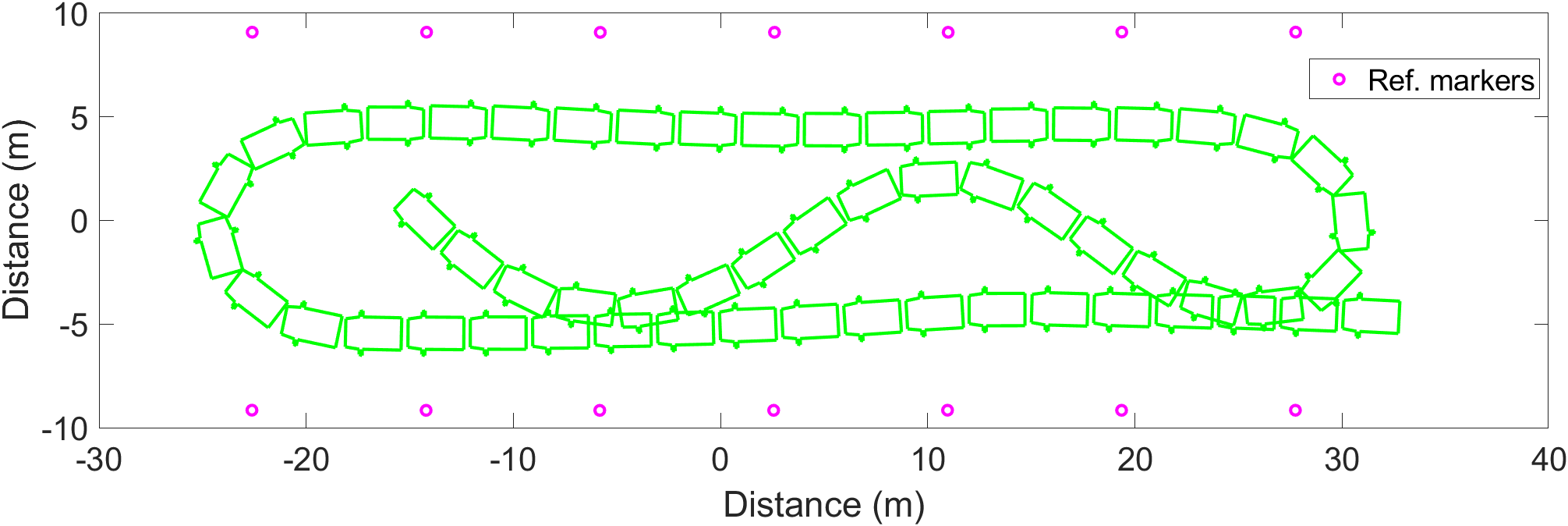}
		\caption{Trajectory of a vehicle (green) moving in an enclosed test facility}
		\label{fig:IndoorNav}
	\end{figure}

	Knowing the performance of the individual modules, the next step is to put them all together to evaluate the performance of the complete algorithm.	For this, a vehicle is driven in an enclosed test facility, and the IMU signals together with the LiDAR measurements are recorded for offline evaluation. The algorithm workflow is as described in \autoref{sec:algorithm}. First, it is detected if the vehicle is standing still. If the vehicle is in motion, its movement is predicted by means of an EKF. Finally, the LiDAR measurements are used as correction data in the EKF. The results are seen in the \autoref{fig:IndoorNav}.
	
	\subsection{Execution Time} \label{sec:Res:Runtime}
	
	Algorithm runtime is one of the most important constraints for autonomous driving. For analysing the possibility of real-time implementation of the proposed algorithm, runtime measurements on an Intel i7-6820HQ CPU are performed. For this, the runtime of 1.1+ million run cycles of each module is measured with the Matlab Profiler. As seen in the results shown in the \autoref{tab:RTresults}, even for high level code, the algorithm is real-time capable.
	
	\ctable[mincapwidth=\columnwidth,
	label=tab:RTresults,doinside=\relscale{0.94},
	caption={Runtime results for the proposed method. Shown are: modules, code type, the median and standard deviation of the execution time on the CPU.}]{lrrr}{}{
		\FL
		Module        &       Code &      Median  &   Std. dev.\LL
		Point clustering  & Matlab &  43 us  & 20 us\NN
		Standstill classifier  & Matlab &  123 us  & 50 us\NN 
		Kalman filter  & Matlab &  403 us  & 98 us\NN 		
		\mc{Velocity estimation} & Matlab & 11 us  & 14 us\NN 
		& Mex &  20 us & 27 us\NN
		\mc{Pose estimation} & Matlab & 42 us & 54 us\NN 
		& Mex &  20 us & 45 us\LL 
	}
	
	\section{Conclusions and Future Work}
	
	In this work, a procedure for a reference vehicle state estimation is presented. The procedure includes a robust, purely IMU-based standstill recognition, a vehicle motion prediction by means of an EKF, and a LbPM that can deliver high-precision correction data in roofed areas (such as enclosed test facilities or parking lots). 
	
	Each of the modules is extensively tested with several hours of real-data and evaluated using a state-of-the-art INS with RTK as ground truth. The results confirm the high accuracy of the proposed method, which closely approximates RTK quality for indoor environments.
	
	The following three factors imply a massive progress for indoor navigation: the achieved high-accuracy, the availability of the proposed procedure where current state-of-the-art sensors are not available and the possibility of real-time implementation.
	
	The future work includes the implementation of the procedure in portable hardware for on-line vehicle state estimation and runtime evaluation in embedded systems. The procedure has improving potential in the area of quality indicators. There are situations where the used motion model does not depict the movement of a vehicle anymore (such as drifting). The automatic generation of quality indicators could function as a self-supervising mechanism to switch between different motion models or fine tuning of the EKF.
	
	\section{Acknowledgement}
	
	The authors acknowledge the financial support by the Federal Ministry of Education and Research of Germany (BMBF) in the framework of FH-Impuls (project number 03FH7I02IA).
	
	\iftemplateconf
	\bibliographystyle{IEEEtran}
	\else
	\bibliographystyle{IEEEtranN}
	\fi
	\bibliography{literature}
	
\end{document}